\documentclass{ifacconf}
\usepackage{microtype}
\usepackage{graphicx}      
\usepackage{natbib}        
\usepackage{subfig}
\usepackage{enumitem}
\usepackage[T1]{fontenc}
\usepackage{dsfont}
\usepackage{url}

\newcommand{\BN}{\mathds{N}}

\newcommand*\TRANS{{\mathpalette\doTRANS\empty}}
\makeatletter
\newcommand*\doTRANS[2]{\raisebox{\depth}{$\m@th#1\intercal$}}
\makeatother

\usepackage{url}
\usepackage{amssymb,mathtools}

\usepackage{dsfont}
\usepackage{tikz}

\newtheorem{remark}{Remark}

\begin{document}
\begin{frontmatter}

\title{Transmission Neural Networks: Approximation and Optimal Control\thanksref{footnoteinfo}} 

\thanks[footnoteinfo]{Supported in part by Simons-Berkeley Research Fellowship (SG), NSERC (Canada) Grants RGPIN-2024-06612 (SG),  RGPIN-2019-05336 (PEC) and US-AFOSR Grant FA9550-23-1-0015 (PEC).}

\author[First]{Shuang Gao} 
\author[Second]{~Peter E. Caines} 

\address[First]{Department of Electrical Engineering, Polytechnique Montreal, Montreal, QC, Canada (e-mail: shuang.gao@polymtl.ca)}
\address[Second]{Department of Electrical and Computer Engineering, McGill University, Montreal, QC, Canada (e-mail: peterc@cim.mcgill.ca)}

\begin{abstract}                
Transmission Neural Networks (TransNNs) introduced by \cite{ShuangPeterTransNN22} connect  virus spread models  over networks and neural networks with tuneable activation functions. 
This paper presents  the approximation technique and  the underlying assumptions employed by TransNNs  in relation to the corresponding Markovian Susceptible-Infected-Susceptible (SIS) model with $2^n$ states, where $n$ is the number of nodes in the  network. The underlying infection paths are assumed to be stochastic with heterogeneous and time-varying transmission probabilities.  
We obtain the conditional probability of infection in the stochastic $2^n$-state SIS epidemic model corresponding to each state configuration  under mild assumptions, which enables control solutions based on Markov decision processes (MDP).
Finally, MDP control with $2^n$-state SIS epidemic models and optimal control with TransNNs are compared in terms of mitigating virus spread over networks through vaccination, and it is shown that TranNNs enable the generation of control laws with significant computational savings, albeit with more conservative control actions. 
\end{abstract}


\end{frontmatter}

\section{Introduction}
Epidemic models play a crucial role in understanding, predicting, and mitigating epidemic spreads, and researchers have proposed and analyzed various epidemic models   \citep{pastor2015epidemic,nowzari2016analysis, kiss2017mathematics, van2014performance,liggett2013stochastic,pare2020modeling}.

Towards modelling the spread of epidemics over networks, 
\cite{lajmanovich1976deterministic} provided a first thorough system theoretic analysis, where the network characterizes transmission probability rates among different groups.  
\cite{kephart1992directed} analyzed  virus spread processes on random directed graphs. Virus spread models on  networks characterized by degree distributions are analyzed  by \cite{pastor2001epidemic}.  %
Mean field approximations  for continous-time Susceptible-Infected-Susceptible (SIS) virus spread models on networks have been established in \citep{van2008virus,cator2012second,van2015accuracy,van2014performance}, where mean field states approximate the fractions of the infected in nodal populations.  Continous-time contact processes can  also model epidemic spreads (see e.g. \citep{liggett2013stochastic}). %

Discrete-time SIS models over networks, which are closely related to the current paper, have been investigated by various researchers.  Discrete-time virus spread models  over networks with homogeneous transmission probabilities have been proposed for identifying threshold values for epidemics spread in \citep{wang2003epidemic,chakrabarti2008epidemic}. 
In \citep{ahn2013global}, the model in  \citep{chakrabarti2008epidemic} is linearized to provide  upper bounds for infection states, and the stability  is then analyzed for the linearized model.  
\cite{ahn2014mixing} further established that the nonlinear discrete-time SIS model in \citep{chakrabarti2008epidemic} provides an upper bound on the probability of infection generated from the discrete-time Markov chain model with $2^n$ states, with $n$ as the number of agents, and such a result was established for the case with homogeneous infection probabilities across all links.
%
%
%
%
Towards extending the discrete-time SIS model in \citep{chakrabarti2008epidemic} to the case with non-homogeneous transmission probabilities, \cite{han2015data}  obtained a linear dynamic model that provides an upper bound for the probability of infection, and then used the linear model to solve vaccine allocation problems via geometric programming. In \citep{paarporn2015epidemic}, 
an   SIS model approximating the $2^n$-state stochastic SIS model over networks was analyzed, which is a different approximation model from the current paper. An observer model was proposed for the discrete-time $2^n$-state stochastic SIS model for designing feedback control  in \citep{watkins2017inference}.
A comparative analysis of two discrete-time SIS epidemic  models, where one is based on the Euler discretization of the continous time SIS model in \cite{van2008virus} and the other is a variant of the discrete time SIS model in \citep{chakrabarti2008epidemic,han2015data,ahn2014mixing}, was carried out by \cite{pare2018analysis}. %
 Both models are shown to converge to the continous time SIS model in \citep{van2008virus} with infinitesimal sampling time (see \cite{pare2018analysis,ShuangPeterTransNN22}). The connection between discrete-time SIS models with non-homogeneous transmission probabilities and neural networks is established by \cite{ShuangPeterTransNN22} via the TransNN model. TransNNs offer  concise representations of SIS epidemic models with heterogenous time-varying transmission probabilities and  networks.

 As  learning models,  TransNNs are universal function approximators and  have trainable activation functions~\citep{ShuangPeterTransNN22,ShuangTransNNVideo22}.    
The key conceptual difference between TransNNs and standard neural networks is that trainable activation functions in TransNNs  are associated with links, whereas in standard neural networks,  
activation functions are typically considered as a part of nodes. {Such a feature with trainable links functions also appeared in  Kolmogorov-Arnold 
networks \citep{liu2024kan}.}

{Main contributions of the current paper are as follows. Firstly, the relation between the  $2^n$-state epidemic model over heterogenous and time-varying transmission probabilities and the TransNN model is established by identifying  necessary technical assumptions.   
 Secondly,  the conditional probability of infection in the stochastic $2^n$-state epidemic model corresponding to each state configuration is established. 
Finally, it is demonstrated that TransNN control formulations can be used to mitigate virus spread over  (time-varying and random) networks and enable significant computational reduction compared to controlling  virus spreads using $2^n$-state stochastic SIS epidemic models. } 
 
\section{Virus Spread Over Networks}
Consider a physical  contact network denoted by $(V,E)$ with an adjacency matrix $A=[a_{ij}]$, where $V$ is the node set and $E\subset V\times V$ is the edge set.  Each node of the physical contact network represents an individual person and a link between two persons exists if they are within a given distance for an extended period of time (e.g. within 2 meters for at least 15mins). 
For all $i,j\in V$, let $w_{ij}^k$ denote the probability of node $j$ infecting its neighbouring node $i$ on the physical contact network given that $j$ is infected at time $k$, and  let $W_{ij}^k \in\{0,1\}$ be the binary random variable representing the successful transmission of virus from node $j$ to node $i$ at  time $k$. For simplicity, let $V =[n]\triangleq \{1,2,..., n\}$.

The state of a node $i \in [n]$ at time $k$ is denoted by a random variable $X_i(k)$, which takes binary values $0$ and $1$, with $0$ representing the healthy state and $1$ representing the infected state.  The one-step update of the  binary random states for epidemic spread over networks given the current states $(X_i(k))_{i\in [n]}$ follows the dynamics
\begin{equation}\label{eq:agent-based}
	1- X_i(k+1) = \prod_{j\in {N}_i^{\circ k}} \Big(1- W_{ij}^kX_j(k)\Big), \quad \forall i \in [n]
\end{equation}
{where $N_i^{\circ k}  \triangleq  \{j: (i,j)\in E^k \} \cup \{i\}$ denotes the neighbourhood set of node $i$ with itself included at time $k$, and $(V, E^k)$ denotes the physical contact graph at time $k$. 
 {The dynamics have the property that for a healthy node to become infected, the infection needs to come from at least one of its neighbours, and furthermore the successful transmission of the virus from  one neighbour of node $i$ is sufficient to infect node $i$.} 
We highlight that the inclusion of  self-loops (with the neighborhood represented by $N_i^{\circ k} $ for agent $i$ at time $k$)  is essential in the characterization of the recovery process  in the virus spread dynamics, as the recovery process is equivalently represented by the self-transmissions with self-loops.
 
Let $X(k) \triangleq [X_1(k)\dots X_n(k)]^\TRANS \in \{0,1\}^n $ and $W^k\triangleq [W_{ij}^k] \in \{0,1\}^{n\times n}$. 
We introduce the following assumptions. 
\begin{description}[style=nextline, leftmargin=1cm, labelindent=0cm]
\item[(A1)] At any time $k> 0$,	 $W^k\triangleq [W_{ij}^k]$ is independent of $\{W^t: 0\leq t < k\}$ and $\{X{(t)}: 0\leq t < k \}$; 
\item[(A2)]  At any time $k\geq 0$, the  binary random variables $\{W_{ij}^k:i,j \in [n]\}$ representing the transmissions are the conditionally independent given the current states $X(k)$. 
\end{description}

The assumption (A1) ensures that transmissions are independent of the history of states and past transmissions, and (A2) ensures that transmissions are conditionally independent given the current infection state.

Under (A1), the dynamics \eqref{eq:agent-based} are Markovian, since 
\begin{equation}
\begin{aligned}
	\mathbb{E}(1- & X_i(k+1)| (X(t))_{t\in [k]}) \\
	& = \mathbb{E}\Big[\prod_{j\in {N}_i^{\circ k}} \big(1- W_{ij}^kX_j(k)\big)| (X(t))_{t\in [k]} \Big]\\
		& = \mathbb{E}\Big[\prod_{j\in {N}_i^{\circ k}} \big(1- W_{ij}^kX_j(k)\big)| X(k)\Big]\\
	& =  \mathbb{E}(1- X_i(k+1)| X(k)).
\end{aligned}
\end{equation} 

Furthermore, under both  (A1) and (A2), %
we have  
\begin{equation*}
\begin{aligned}
\mathbb{E}(1- X_i(k+1)|X(k)) & = \mathbb{E}\prod_{j\in {N}_i^{\circ k} } \big(1- W_{ij}^kX_j(k)|X(k)\big) \\
& = \prod_{j\in {N}_i^{\circ k} }\mathbb{E}  \big(1- W_{ij}^kX_j(k)|X(k)\big)
\end{aligned}
\end{equation*} 
which corresponds to the virus spread model with one-step update proposed in \citep{ShuangPeterTransNN22}, since  the conditional probability satisfies
 \begin{equation}
1-	\bar{\rho}_i(k+1) = \prod_{j\in {N}_i^{\circ k}} \Big(1- w_{ij}^k\bar{\rho}_j(k)\Big), \quad i\in [n]
\end{equation}
where  
$
w_{ij}^k\triangleq \text{Pr}(W_{ij}^k=1 {| X_j(k)=1})
$ and  $$\bar{\rho}_i(k)\triangleq \mathbb{E}(X_i(k)|X(k-1)) = \text{Pr}(X_i(k)=1|X(k-1)).$$ 
For a state configuration $q\in\{0,1\}^n$, 
\begin{equation*} 
\begin{aligned}
	\text{Pr}(&X(k+1)=q|X(k))   = \prod_{i=1}^n \text{Pr}(X_i(k+1)=q_i|X(k)) 
\end{aligned}
\end{equation*}
where the equality is due to the conditional independence of the virus transmissions $\{ W_{ij}^k\}$ assumed in (A2). 

\begin{prop}[Conditional Probability of Infection]~~~\\
Assume (A1) and (A2) hold. 
Given the state configuration at time $k$ denoted by $X(k)= x(k) \in \{0,1\}^n$, the transition probability to a state configuration $q\in\{0,1\}^n$ is  given by 
\begin{equation}\label{eq:MarkovTranProb2}
\begin{aligned}
	\text{Pr}(X(&k+1)=q|X(k)=x(k))\\ & = \prod_{i=1}^n \text{Pr}(X_i(k+1)=q_i|X(k)= x(k))\\
	& = \prod_{i=1}^n \Big(q_i\rho_i(k+1) + (1-q_i) (1-\rho_i(k+1)) \Big)
\end{aligned}
\end{equation}
where
 \begin{equation}\label{eq:prob-configuration}
\rho_i(k+1) = 1-	\prod_{j\in {N}_i^{\circ k}} \Big(1- w_{ij}^k x_j(k)\Big), \quad i\in [n].
\end{equation}
\end{prop}

This explicit representation of the transition probability to the next state configuration $q$ given the current state configuration $x(k)$ will be used later in the Markov Decision Process (MDP) for mitigating virus spread over networks in Section \ref{sec:control-virus-spread}. 

\section{Approximation by TransNNs}
To derive TransNNs that approximate the virus spread dynamics \eqref{eq:agent-based}, we  introduce the following  assumptions.  %
\begin{description}[style=nextline, leftmargin=1cm, labelindent=0cm]
	\item[(A3)]  At time $k\geq 0 $, $\{ W_{ij}^k: i,j \in [n]\}$ are  independent and $W_{ij}^k$ is independent of  $\{ X_i(k): i\in [n],  i\neq j\}$ for all $j \in [n]$. 
	\item[(A4)] The  events $\{\{X_i(k)=1\}: i\in [n], 0\leq k\leq T \}$ are independent with $T$ as the terminal time. 
\end{description}
The assumption (A3) is stronger than (A2) and introduces the independence of transmissions over different links at the current time $k$ and the independence of transmissions with respect to the states except the current state at the sending node.  

Assumptions (A3) and (A4)  allow us to write the update of the expected state by taking the (total) expectation inside the product on the right-hand side  as follows: for $i\in [n]$, 
\begin{equation}\label{eq:naive-mf-approx}
	\begin{aligned}
\mathbb{E}(1- X_i(k+1)) & = \mathbb{E}\prod_{j\in {N}_i^{\circ k} } \big(1- W_{ij}^kX_j(k)\big) \\
& = \prod_{j\in {N}_i^{\circ k} }\mathbb{E}  \big(1- W_{ij}^kX_j(k)\big) .
\end{aligned}
\end{equation}
Under   (A3) and (A4), the joint probability distribution for the random variables $\{Z_{ij}(k )\triangleq 1- W_{ij}^kX_j(k), {j\in {N}_i^{\circ k} } \}$ on the right-hand side is equal to the product of its marginal distributions.  
If (A3) and (A4) are not satisfied, the right-hand side of \eqref{eq:naive-mf-approx} constitutes an approximation of the joint distribution by the product of the marginals,  referred to as the naive mean-field approximation \citep{barber2012bayesian} or individual-based mean-field approximation  \citep{van2008virus}. 

Let $w_{ij}^k\triangleq \operatorname{Pr}(W_{ij}^k=1 {| X_j(k)=1})$ denote the conditional probability of the successful transmission of the virus from node $j$ to node~$i$ at time step $k$. 
Then
\[
\mathbb{E}  \big(1- W_{ij}^kX_j(k)\big) =  \big(1-   w_{ij}^k \mathbb{E}  X_j(k)\big), \quad \forall i, j \in [n]. 
\] 
Hence under (A1),  (A3) and (A4), the dynamics in \eqref{eq:naive-mf-approx} become the  virus spread model in \cite{ShuangPeterTransNN22}:
\begin{equation}\label{eq:TransNNs-virus}
1-	p_i(k+1) = \prod_{j\in {N}_i^{\circ k}} \Big(1- w_{ij}^kp_j(k)\Big), \quad i\in [n],
\end{equation}
with  $p_i(k)\triangleq \text{Pr}(X_i(k)=1) = \mathbb{E}X_i(k) $ representing the (total) probability of node $i$ being infected at time $k$.
Furthermore, via the nonlinear state transformation     
\[ 
	s_i(k)\triangleq -\log(1-p_i(k)) ~ \in [0,\infty],
\]  
proposed by \cite{ShuangPeterTransNN22},
the equivalent TransNN representation is then given by  
\begin{equation}\label{eq:TransNN}
	s_i(k+1)=\sum_{j\in {N}_i^{\circ k}} \Psi(w_{ij}^k, s_j(k)), \quad i \in [n]
\end{equation}
where
\begin{equation}\label{eq:TlogSigmoid}
\Psi(w,x)= -\log(1-w+we^{-x})	
\end{equation}
is  the TlogSigmoid activation function in \citep{ShuangPeterTransNN22}
 with the activation level $w\in[0,1]$ and the input signal $x\in[0,+\infty]$. Interested readers are referred to  \citep{ShuangPeterTransNN22,ShuangTransNNVideo22} for detailed properties of the TransNN model and its connections with standard neural network models. 

 Thus, in  the stochastic model with $2^n$ states described in \eqref{eq:agent-based}, imposing the independence assumptions (A3) and (A4) together with (A1)  leads to the virus spread model in  \eqref{eq:TransNNs-virus} and the equivalent TransNN model  \eqref{eq:TransNN}. 

\begin{remark}[Discussions on Assumption (A4)]
{In general the assumption (A4) may not hold depending on the actual paths of infection over the networks.}
There are two ways to get around the assumption on independence in (A4). One way to break the potential dependence among  $\{X_i(k)=1,i\in [n]\}$ is to reset or observe the states at each time step (i.e. $T=1$). The disadvantage of this approach is that we need to observe or simulate the states at each time step.  A similar idea is  used in the context of Restricted Boltzmann Machines (RBMs) \citep{salakhutdinov2007restricted} if we interpret the time steps in the current paper as layers of bipartite graphs in RBMs. 
Alternatively,  we can sample points from the joint probability distribution to approximate and simulate the evolutions of the empirical distributions. The disadvantage lies in the need to keep track of the evolution of the (empirical) joint distribution over the state space with $2^n$ possible configurations at each time.

Both approaches may lead to significant computational burden especially when the underlying network is large.
Instead, one can still explore the approximated model in \eqref{eq:TransNNs-virus} to have an approximate characterization of the probability of infection with low complexity, {given only the infection status (or probability) at the initial time}. 	 
\end{remark}	 

In the following, we show that without the assumption (A4) the  dynamics in \eqref{eq:TransNNs-virus} (and equivalently \eqref{eq:TransNN}) provide an upper bound for the infection probability given by the stochastic virus spread model in \eqref{eq:agent-based}. 

\begin{prop}[\cite{ShuangPeterTransNNGERAD25}]\label{prop:prob-upper-bound}
Assume (A1) and (A3) hold. 
Let $X_i(k)$ be the binary state in dynamics \eqref{eq:agent-based} and let $p_i(k)$ denote the probability state in  dynamics \eqref{eq:TransNNs-virus}. Given the same physical contact network $(V, E^k)$ for all  $k \in \mathds{N}$, the same conditional transmission probabilities $\{w_{ij}^k, i,j \in [n], k\geq 0\}$ and the same initial infection probability (i.e. $p_i(0)=\operatorname{Pr}(X_i(0)=1)$ for all $i\in [n]$), the following inequality holds 
\begin{equation}\label{eq:prop1}
	p_i(k) \geq \operatorname{Pr}(X_i(k)=1), \quad \forall i \in [n], ~~\forall k \in \BN;
\end{equation}
if, furthermore, \textup{(A4)} is satisfied for the dynamics \eqref{eq:agent-based},  then the equality holds for all $k\in\{0,1,\dots, T\}$. 
\end{prop}

A similar inequality holds for Shannon information state (of being healthy). 
Consider the following state transformation \citep{ShuangPeterTransNN22}:
$$
\mathbb{T}(p)= -\log(1-p) \in [0,\infty],\quad  \forall p\in[0,1].
$$
The state transformation above is bijective (from $[0,1]$ to $[0, \infty]$) and monotonically increasing with respect to the input. Taking $\mathbb{T}(\mathbb{E}X_i)$ as the state of node $i$, then the dynamics in \eqref{eq:update-inequality}
 \begin{equation}\label{eq:update-inequality}
 	 \mathbb{E}X_i(k+1) \leq 1- \prod_{j\in {N}_i^{\circ k}} \Big(1- w_{_{ij}} \mathbb{E}X_j(k)\Big),\quad \forall i \in [n]
 \end{equation}
 leads to the following upper bound dynamics
\[
\mathbb{T}(\mathbb{E}X_i(k+1))\leq \sum_{j\in N_i^{\circ k}}  \Psi(w_{ij}, \mathbb{T}(\mathbb{E}X_i(k)))
\]
in terms of the evolution of the Shannon information state (of being healthy).
Then by Proposition~\ref{prop:prob-upper-bound}, we obtain that
\begin{equation}
	s_i(k)\geq \mathbb{T}(\mathbb{E}X_i)\triangleq -\log(1-\mathbb{E}X_i),
\end{equation}
that is, the TransNN dynamics in \eqref{eq:TransNN} provide an upper bound for the information content of being healthy. 
\begin{remark}
	{Under (A1) and (A3), the  $2^n$-state Markovian SIS dynamics  \eqref{eq:agent-based} have an absorbing state (which is the state where all nodes are healthy). It implies that when the time runs sufficiently long, all nodes will eventually become healthy. However, when $n$ is large, the time required for the infection to end can be extremely long, and in such cases the expected mixing time can be analyzed instead which depends on the network connectivity \citep{ahn2014mixing}.}
\end{remark}

\begin{prop}
	Let $X_i(k)$ be the binary state in the dynamics \eqref{eq:agent-based}. Assume  \textup{(A1)} and \textup{(A3)} hold. Let $\Omega_k \triangleq  [w_{ij}^k]$ denote the matrix of the conditional probabilities of transmissions. Then for all $i \in [n]
$, the following holds
	\[
	 \operatorname{Pr}(X_i(k)=1) \leq [(A_k\odot \Omega_k)  \dots (A_0\odot \Omega_0) \mu(0)]_i 
	\]
	where the initial condition is  $\mu(0) \triangleq [\mu_1(0),\dots, \mu_n(0)]$, $\mu_i(0) \triangleq \operatorname{Pr}(X_i(0)=1)$, and $\odot$ is the Hadamard product. 
\end{prop}
\begin{pf}
		An application of Prop. \ref{prop:prob-upper-bound} and  the proof argument in  \cite[Thm. 1]{ShuangPeterTransNN22} that shows the TransNN model is less conservative than the linear system approximation of the SIS epidemic model \eqref{eq:TransNNs-virus} lead to the desired result.
\end{pf}

\section{Controlling Virus Spread on Networks} \label{sec:control-virus-spread}
Let the control $u_i(k) \in\{0,1\} $  represent the effort (e.g.  vaccination) at node $i$ to reduce individual probability of getting infection; more specifically, at time $k$, $u_i(k) =1$ if individual $i$ gets vaccinated and $u_i(k)=0$ otherwise.  
Assume the vaccination  reduces $1- \beta$ of the infection probability. For a control action $u(k) \triangleq (u_1(k)\dots u_n(k))^\TRANS$, the controlled transmission probabilities are then given by 
\vspace{0.1cm}
\begin{equation} \label{eq:controlled-probability}
m_{ij}^{k}(u_i(k)) \triangleq u_i(k) w_{ij}^k \beta + (1-u_i(k)) w_{ij}^k, ~ \forall i, j \in  [n].
\end{equation}

In the following, we first introduce the MDP-based solutions, followed by the control solutions using TransNNs. The two solutions will then be compared via simulations.

\subsection{Optimal MDP Solutions with  Markovian SIS Dynamics}
In the MDP formulation with the $2^n$-state stochastic dynamics in \eqref{eq:agent-based},  consider the cost
\begin{equation}\label{eq:J1}
	J_1 = \mathbb{E}\sum_{k=0}^{T-1} \mathbf{1}_n^\TRANS( c X(k) +   u(k))  \triangleq \mathbb{E}\sum_{k=0}^{T-1}  l (X(k), u(k))
\end{equation}
with transition probabilities given by \eqref{eq:MarkovTranProb2}. In this case, the state space is $\{0,1\}^n$ and the control action space is $\{0,1\}^n$. 
Let $V_k: \{0,1\}^n \to \mathbb{R}$ denote the value function defined below: 
$$
V_k(x) \triangleq \min_{u_{k}\dots u_{T-1}} \mathbb{E}\sum_{t=k}^{T-1}  l (X(t), u(t)), \quad X(k) = x. 
$$
By Dynamic Programming (see e.g. \citep{bertsekas2017dynamic}), the value function satisfies the Bellman equation
\[
V_k(x) = \min_{u} \bigg[ l(x, u) + \sum_{x^\prime \in \{0,1\}^n}\text{Pr}(x^\prime|x, u ) V_{k+1}(x^\prime) \bigg] 
\]
where $V_T =0$,  the transition probability is specified by 
\begin{equation}
\begin{aligned}
	\text{Pr}(&X(k+1)=q|X(k)=x(k), u(k)=u(k))\\
	& = \prod_{i=1}^n \Big(q_i\rho_i(k+1) + (1-q_i) (1-\rho_i(k+1)) \Big)
\end{aligned}
\end{equation}
with
$
\rho_i(k+1) = 1-	\prod_{j\in {N}_i^{\circ k}} \Big(1- m_{ij}^{k}(u_i(k))x_j(k)\Big)
$
and $m_{ij}^{k}(u_i(k))$ in \eqref{eq:controlled-probability}. The optimal control corresponds to the minimizer of the right-hand side of the Bellman equation.

\subsection{Optimal Control based on TransNNs}
Consider the same cost (adapted from $J_1$\footnote{We note that $J_1 = J_2$ if the assumption (A4) holds, and in general $J_1\leq  J_2$ holds if the control actions are the same.} in \eqref{eq:J1})
\[
J_2 = \sum_{k=0}^{T-1}\sum_{i=1}^n ( c p_i(k) + u_i(k)) = \sum_{k=0}^{T-1} \mathbf{1}_n^\TRANS (c p(k) +  u(k))
\]
with the dynamics given by \eqref{eq:TransNNs-virus}. %
Equivalently, we can use the TransNN model \eqref{eq:TransNN} for the dyamics
\begin{equation}
	s_i(k+1)=\sum_{j\in {N}_i^{\circ k}} \Psi(m_{ij}^{k}(u_i(k)), s_j(k)), \quad i \in [n],
\end{equation}
with $
\Psi(w,x)= -\log(1-w+we^{-x})	
$
 and the cost is then equivalently given by 
\[
J_2 = \sum_{k=0}^{T-1} \mathbf{1}_n^\TRANS (c (\mathbf{1}_n - \text{exp}_{\circ}(-s(k))) +  u(k)),
\]  
where $\text{exp}_\circ$ denotes the element-wise exponential function.  The state space is $[0,\infty]^n$ and the space of control actions  is $\{0,1\}^n$. The standard Minimum Principle (see e.g. \citep{bertsekas2017dynamic}) does not directly apply to the optimal control problem above with binary control actions as the needle variation technique in proving the Minimum Principle does not hold. 
To apply the Minimum Principle, we need to reformulate the optimal control problem with relaxed control actions in $[0,1]^n$ (i.e. $u_i(k) \in [0,1]$ for all $i \in [n]$ and for all $k\geq 0$). For the relaxed control problem,  
the Hamiltonian is given by
\begin{equation}\label{eq:Hamiltonian}
\begin{aligned}
	H(k) = & ~\mathbf{1}_n^\TRANS \left( c (\mathbf{1}_n - \exp_{\circ}(-s(k))) + u(k) \right) \\
	& + \sum_{i=1}^{n} \lambda_i(k+1) \sum_{j\in {N}_i^{\circ k}} \Psi(m_{ij}^{k}(u_i(k)), s_j(k)).
\end{aligned}
\end{equation}
The adjoint dynamics is then given by 
\[
\begin{aligned}
	\lambda_i(k)  
	& = c  e^{-s_i(k)}  \\& +  \sum_{\ell \in \mathcal{N}_{i-}^{\circ k} } \lambda_\ell(k+1)    \frac{\partial}{\partial s_i}   \Psi(m_{\ell i}^{k}(u_\ell(k)), s_i(k)) 
\end{aligned}
\]
with $ \lambda_i(T) = 0$, where  $\mathcal{N}_{i-}^{\circ k}: =\{\ell: (\ell, i)\in E^k\} \cup i$ denotes the outgoing neighborhood of $i$ with itself included, and $m_{ij}^{k}(u_i(k))$ given by \eqref{eq:controlled-probability}  and 
\[
\begin{aligned}
	\frac{\partial}{\partial s_i}  \Psi( m_{\ell i}^{k}&(u_\ell(k)), s_i(k))  =\\
	& \frac{m_{\ell i}^k (u_\ell(k)) e^{-s_i(k)}}{1 - m_{\ell i}^k (u_\ell (k))  + m_{\ell i}^k (u_\ell (k))  e^{-s_i(k)}} .
\end{aligned}
\]

One can check the optimal control action  node by node  without loss of generality since control actions of different nodes affect the Hamiltonian in \eqref{eq:Hamiltonian} in a decoupled way.
Moreover, it is easy to verify that the Hamiltonian $H(k)$ in \eqref{eq:Hamiltonian} is convex in $u_i(k)$ for all $i\in [n]$ and $k\geq 0$. Therefore the optimal relaxed  control action exists and lies either on the boundary or in the interior.

Since the original control action  $u_i(k) \in \{0,1\}$ is binary, an (approximate) optimal control is determined by 
\begin{equation}
\Delta H_i(k) = H(k) |_{u_i(k)=1} - H(k) |_{u_i(k)=0}
\end{equation}
which is explicitly given by
\begin{equation*}
\Delta H_i(k) = 1 - \sum_{j\in {N}_i^{\circ k}} \lambda_j(k+1) \log \frac{1 - w_{ij}^k \beta + w_{ij}^k \beta e^{-s_j(k)}}{1 - w_{ij}^k + w_{ij}^k e^{-s_j(k)}}.
\end{equation*}
One can verify that (i) the last term in the equation above is negative and (ii) the adjoint process is non-negative based on its dynamics. Thus, $ \Delta H_i(k)$ could be either positive or negative.
Then the (approximate) optimal control rule is  given by
\begin{equation} \label{eq:optimal-control}
u_i^*(k) =
\begin{cases}
1, &   \Delta H_i(k)  <  0  \\
0, & \text{otherwise}.
\end{cases}
\end{equation}
 To verify that the control actions generated using \eqref{eq:optimal-control} actually minimize the Hamiltonian, we compare the gradient evaluations $\frac{\partial H(k)}{\partial u_i(k)}|_{u_i(k)=0}$ and $\frac{\partial H(k)}{\partial u_i(k)}|_{u_i(k)=1}$ along the trajectories of states and adjoint states generated using the control \eqref{eq:optimal-control}: if both gradients share the same sign, the convexity of the Hamiltonian $H(k)$ in  $u_i(k)$ implies that the control action that minimizes the Hamiltonian lies on the boundary and hence given by \eqref{eq:optimal-control}.
 
Since the Minimum Principle only provides necessary conditions for optimality,  the optimal control generated above only provides candidates of optimal control solutions. To demonstrate that the solution is globally optimal, additional verification procedures are required.

\subsection{Numerical Results}
The parameters are: $\beta =0.3, ~T=10, ~c=100.$ For the clarity of illustrations, we select $n = 5$. A much larger $n$ is possible for  TransNN-based control, but not feasible for MDP solutions. The network structure and transmission probabilities are shown in Fig.~\ref{fig:net-prob}. 
\begin{figure}[htb] 
\centering
	\includegraphics[width=7.5cm]{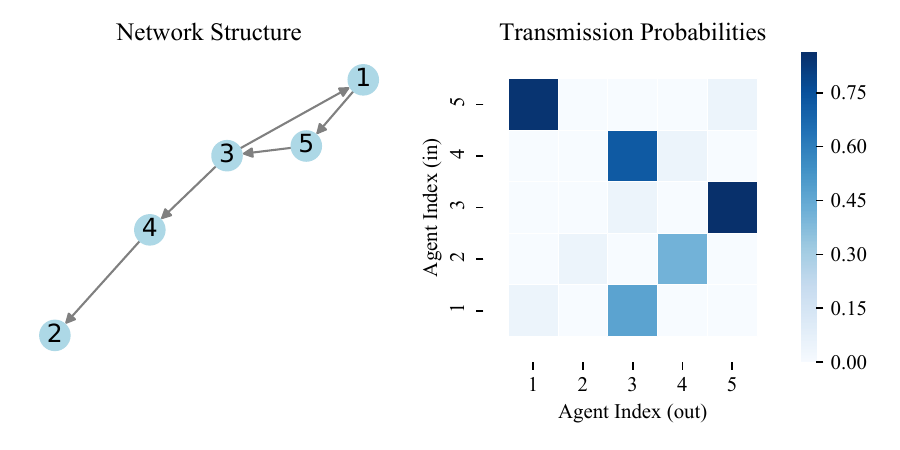}
	\caption{A network example with 5 agents (left) and  transmission probabilities (right) among nodes.} \label{fig:net-prob}
\end{figure}
\begin{figure}[htb]
\centering
		\includegraphics[width=8cm]{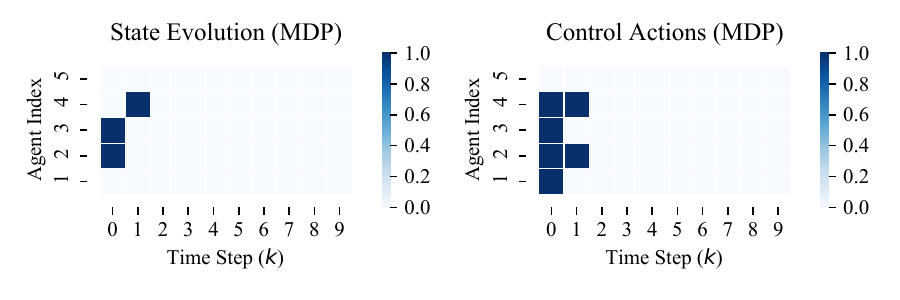}
		\caption{Control actions (right) generated from solving the MDP problems with $2^n$ states and $2^n$ control and the actual states (left) the under such control actions.}\label{fig:mdp-results}
\end{figure}
\begin{figure}[htb]
\centering
		\includegraphics[width=8cm]{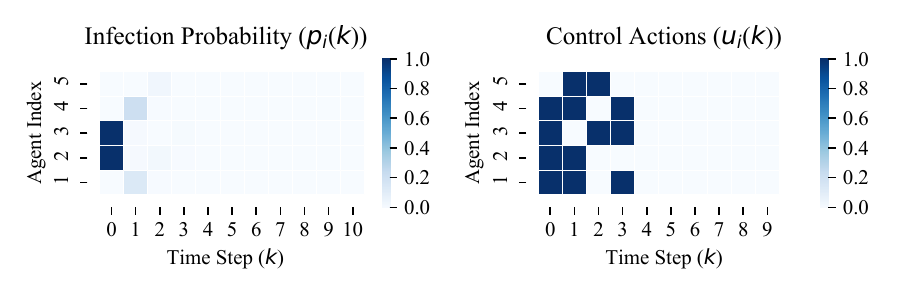}
		\caption{Control actions (right) generated from solving the optimal control problem characterized by TransNN and the infection probabilities (left) in the TransNN model under such control actions.} 
		\label{fig:trans-control-results}
\end{figure}
\begin{figure}[htb]
\centering
		\includegraphics[width=8cm]{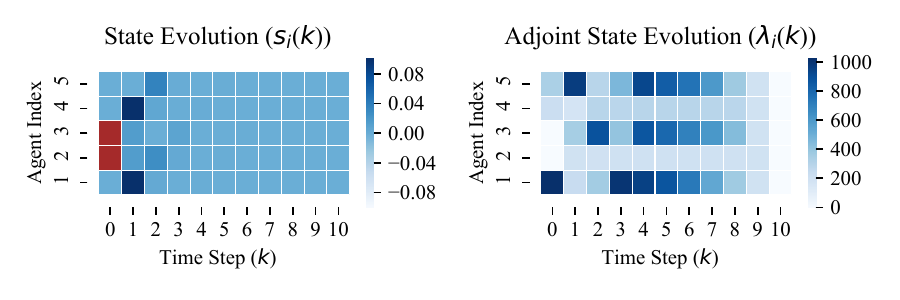}
		\caption{States of TransNNs (left)  under optimal control actions and the adjoint states (right). A brown square (left) represents $s_i(0)= \infty$ (i.e.  $p_i(0)=1$).}
		\label{fig:info-adjoint-states}
\end{figure}
\begin{figure}[htb]
	\subfloat[Computation time  w.r.t. the time horizon $T$.]{
		\includegraphics[width=4cm]{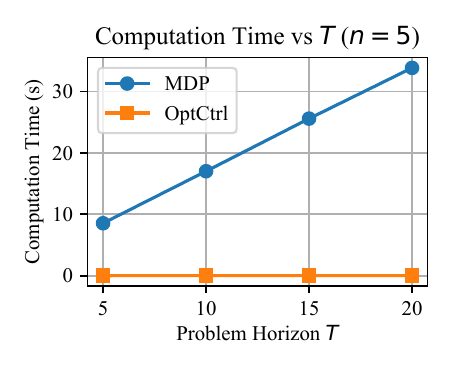}\label{fig:compute_time_vs_T}}
		\quad 
	\subfloat[Computation time w.r.t. the number of agents $n$. ]{\includegraphics[width=4cm]{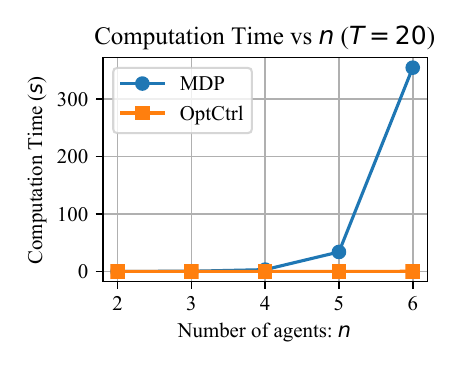}
			\label{fig:compute_time_vs_n}	}
\caption{The computation time for both (i) MDP solution based on Markovian SIS dynamics and  (ii) approximate optimal control (OptCtrl) based on TransNNs. } \label{fig:compute_time}
\end{figure}

The state realizations and the control actions in one simulation under the optimal control from solving the MDP is given in Fig.~\ref{fig:mdp-results}. The execution time for solving the MDP problem is 16.191 seconds on a standard MacBookPro laptop.  
 The probability states of TransNNs and the control actions generated from controlling TransNNs are shown in Fig.~\ref{fig:trans-control-results}, and the adjoint states and the (information) states of TransNNs  are shown in Fig.~\ref{fig:info-adjoint-states}.  The execution time for solving the TransNN-based control problem is 0.020 seconds on the same laptop (which corresponds to a computational time reduction by about 3 orders of magnitude compared to solving MDP).   In Fig.~\ref{fig:compute_time}, the computational time for solving the TransNN-based  control problems and that for solving the MDP problem are illustrated with different choices of $n$ and $T$. 
 From Fig.~\ref{fig:mdp-results} and Fig.~\ref{fig:trans-control-results}, we see that TransNN-based control actions include all the control actions generated from MDP solutions,  and at the first time step the control actions under both solution methods are the same. 

\section{Conclusion}
This work presents the approximation used in TransNNs and compares TransNNs with the associated Markovian SIS model with $2^n$ states for virus spreads.  
 Control formulations based on TransNNs to mitigate virus spread over heterogenous networks enables significant computational reduction compared to solving the MDP problem for virus spreads using $2^n$-state epidemic models.  
 
Future work will (a) evaluate the receding horizon control for TransNNs and control problems with other types of control actions, (b) identify  TransNN models with immune or inhibition states,  (c) relax the condition (A1) to investigate the case with non-Markovian dynamics,  (d) consider the cases with risk-aware control cost, and (f) investigate similar problems in the game-theoretical context where each individual agent has its own objective function.

\begin{ack}
SG would like to thank  Christian Borgs, George Cantwell, Jorge Velasco-Hern\'andez, Xin Guo, Minyi Huang and Aditya Mahajan  for helpful feedback, and Charlotte Govignon for discussions leading to Prop. 1. 
\end{ack}

\bibliography{/Users/smartgao/Dropbox/ImportantBib/mybib}
\end{document}